\documentclass[aps,showpacs,showkeys,preprintnumbers,amsmath,amssymb]{revtex4}
\usepackage{amsmath,amssymb,amsthm}
\usepackage{color}
\usepackage{epsfig}
\usepackage{multirow}
\usepackage{graphicx}
\usepackage{subfigure}
\usepackage{rotating}

\newcommand{\dif}{\textrm{d}}
\newcommand{\Mpl}{\textrm{M}_{\textrm{pl}}^2}
\newcommand{\Lam}{\Lambda^{4}}

\begin{document}
\title{Strong gravitational lensing for black hole with scalar charge in massive gravity}
\author{Ruanjing Zhang, Jiliang Jing\footnote{Corresponding author, Email: jljing@hunn.edu.cn} and Songbai Chen }
\affiliation{Department of Physics, Key Laboratory of Low Dimensional Quantum Structures and Quantum Control of Ministry of Education, and Synergetic Innovation Center for Quantum Effects and Applications, Hunan Normal University, Changsha, Hunan 410081, P. R. China
}

\begin{abstract}
We investigate the strong gravitational lensing for black hole with scalar charge in massive gravity. We find that the scalar charge and the type of the black hole significantly affect the radius of the photon sphere, deflection angle, angular image position, angular image separation, relative magnifications and time delay in strong  gravitational lensing. Our results can be reduced to that of the Schwarzschild  and  Reissner-Nordstr$\ddot{o}$m black holes in some special cases.

\end{abstract}


\maketitle

\section{Introduction}

From the general relativity (GR), we know that photons would be deviated from their straight path when they pass close to a compact and massive body, and this deflection of light was first observed in 1919 by Dyson, Eddington and Davidson \cite{Dyson}. The effect resulting from the deflection of light rays in a gravitational field is known as gravitational lensing and the object causing a detectable deflection is usually named a gravitational lens \cite{Einstein}. Gravitational lensing can help us extract the information about the distant stars which are too dim to be observed, similar as a natural and large telescope. An astrophysical object (such as black hole) makes light rays passing close to it to have a large deviation, even makes a complete loops around the object before reaching the observer, resulting in two infinite sets of the denominated relativistic images on each side of the object. We could not only extract the information about black holes in the universe, but also verify profoundly alternative theories of gravity in their strong field regime with these relativistic images \cite{Vir,Vir1,Vir2,Vir3,Fritt,Bozza,Bozza2,Eirc,Eirc1,whisk,Gyulchev,Bhad1,TSa1,AnAv,gr1,Kraniotis,JH}. Therefore, the strong gravitational lensing is regarded as a powerful indicator of the physical nature of the central celestial objects and then has been studied extensively in various theories of gravity \cite{schen} recent years.

Several theories \cite{Brans,Buchdahl,Gia,Jacob,Ferraro,Kezban} were proposed to generalize GR in order to get an agreement with the observations that the universe is going through a phase of the accelerated expansion \cite{Adam,S} without requiring the existence of the cosmological constant or dark energy and dark matter. There is a widely know model which is a well-developed case in the infrared modification of gravity and all of these points are nicely illustrated, called massive gravity \cite{Dubovsky}. Fierz and Pauli \cite{Pauli} did the first attempt to include mass for the graviton in 1939. However, the massive gravity theory was not concerned until the vigorous development of quantum field theory in the early 1970s.
Since then, many significant massive gravity models as a modified Einstein gravity theory (see the reviews on the subject \cite{Rham,Kurt,Michael}) were proposed, especially in recent years. In this paper, we focus on a static vacuum spherically symmetric solution in massive gravity raised by \cite{Michael1,Comelli}. There are many people that pay their attention to this solution. Sharmanthie Fernando \cite{Sharmanthie,Sharmanthie1} studied quasinormal modes of scalar and massless Dirac perturbations of this theory. Then, Fabio Capela \cite{Fabio1,Fabio} checked the validity of the laws of thermodynamics in massive gravity by making use of the exact black hole solution and equilibrium states and phase structures of such a solution enclosed in a spherical surface kept at a fixed temperature. Furthermore, the geodesic structure of this solution was discussed in Ref. \cite{zhang}. In this paper, we will study the strong gravitational lensing of this solution.

Scalar fields obtained much attention during the recent years. There are several reasons for this, first and foremost, scalar fields both as fundamental fields and effective fields, are well motivated by the standard model particle physics. Then, we want to explore different field contents, in order to check whether the ``No Hair Theorem" \cite{Ruffini} is true and explore the structure of black holes. Scalar fields are one of the simplest types of ``matter" which are often considered by physicists. Finally, the presence of the scalar fields lead to different black hole spacetimes than those of GR, and may engender some new phenomena. We hope one could detect these deviations in astrophysical observations. Moreover, G. Aad \cite{Aad,Chatrchyan} et al. discovered a scalar particle at the Large Hadron Collider, at CERN, which is identified as the standard model Higgs boson since 2012. This observational proves that there is fundamental scalar fields existing in nature. And many examples of scalar hairy black holes \cite{Martinez,Nadalini,Anabalon,Herdeiro} were constructed by passing the long standing ``No Hair Theorem". Therefore, it is important to study the strong gravitational lensing for black hole with scalar hair. This paper shows that the scalar hair has great influence on the strong gravitational lensing.

The outline of this paper is as follows. In Sec.II we study the physical properties of the strong gravitational lensing around the black hole with scalar charge in massive gravity and probe the effects of the deformation parameters on the radius of the photon sphere, minimum impact parameter and deflection angle. In Sec.III we suppose that the gravitational field of the supermassive black hole at the center of our galaxy can be described by this metric and then obtain the numerical results for the main observables in the strong gravitational lensing, such as the angular image position, angular image separation and relative magnifications of these images. Then, we obtain the time delay of light traveling from the source to the observer by numerical calculation in Sec.IV. We end the paper with a summary.

\section{Deflection angle in massive gravity}

We will use the following massive gravity model \cite{Dubovsky:2004sg}
\begin{eqnarray}
\label{eq:action}
\mathcal{S} = \int \dif x^{4} \sqrt{- g}
\left[ - \Mpl \mathcal{R} + \mathcal{L}_{\textrm{m}} +
\Lam \mathcal{F} \right] ,
\end{eqnarray}
where the first two terms are the curvature and Lagrangian of the minimally coupled ordinary matter which comprise the standard GR action, and the third term describes four scalar fields $\phi^0$, $\phi^i$
whose space-time dependent vacuum expectation values break
spontaneously the Lorentz symmetry. The function ${\cal F}$
depends on two particular combinations of the derivatives of the
Goldstone fields, $\mathcal{F} = \mathcal{F} \left( X, W^{ij}
\right)$, with
\begin{eqnarray*}
X &=& \dfrac{\partial^{\mu} \phi^0 \partial_{\mu} \phi^0}{\Lambda^4} ,~~
W^{ij} =\dfrac{\partial^{\mu} \phi^i
\partial_{\mu}\phi^j}{\Lambda^4}
- \dfrac{\partial^{\mu} \phi^i \partial_{\mu}\phi^0\,
\partial^{\nu} \phi^j \partial_{\nu}\phi^0}{\Lambda^8 X} ,
\end{eqnarray*}
where the constant $\Lambda$ has the dimension of mass.
A static spherically symmetric solution in massive gravity is described by \cite{Michael1,Comelli}
\begin{equation}\label{metric1}
ds^2=-f(r)dt^{2}+\frac{dr^{2}}{f(r)}+r^{2}(d\theta^{2}+\sin^{2}\theta d\varphi^{2}),
\end{equation}
with
\begin{equation}\label{metric2}
f(r)=1-\frac{2M}{r}-\frac{S}{r^{\lambda}},
\end{equation}
where $M$ accounts for the gravitational mass of the body, $\lambda$ is a parameter of model which  depends on the potential, $S$ is a scalar charge whose presence reflects the modification of the gravitational interaction as compared to GR, and we will get standard Schwarzschild solution when $S=0$. The solution (\ref{metric1}) has an attractive behavior at large distances with positive $M$. However, the corresponding Newton potential is repulsive at large distances and attractive near the horizon with negative $M$. The latter does not have a corresponding case in GR, so we only consider the case $M>0$. For $S>0$, the modified black hole has attractive gravitational potential at all distances and the event horizon size is larger than $2 M$. For $S<0$, the event horizon exists only for sufficiently small $S$. But when the event horizon exists, the gravitational field is attractive all the way to the horizon while the attraction is weaker than in the case of the usual Schwarzschild black hole, and the horizon size is smaller. Fortunately the massive gravity theory with a Lorentz violation gives us an asymptotically flat spherically symmetric space with finite total energy, featuring an asymptotic behavior slower than $1/r$ and generically of the form $1/r^{\lambda}$, which makes the black hole solution be far richer than in GR due to the presence of ``hair   $\lambda$". The solution does not describe asymptotically flat space when $\lambda<0$ and the Arnowitt-Deser-Misner(ADM) mass will be infinite when $0<\lambda<1$. For $\lambda>1$, the solution recovers standard Schwarzschild term at large distances and the ADM mass is equal to $M$. So we will limit ourselves to the case $\lambda>1$ in the following. It should note that the form of this metric is the same as Reissner-Nordstr$\ddot{o}$m metric when $\lambda=2$ and $S<0$.

To study the gravitational lensing, we just consider the equatorial plan ($\theta=\frac{\pi}{2}$) as usual. It means that both the observer and the source lie in the equatorial plane and the whole trajectory of the photon is limited on the same plane. For simplicity, we set $r/2M=r$ and $S/(2M)^{\lambda}=S$ in the following calculations. The metric (\ref{metric1}) can be rewritten as
\begin{equation}\label{metric3}
ds^2=-A(r)dt^2+B(r)dr^2+C(r)d\phi^2,
\end{equation}
with
\begin{equation}\label{metric4}
A(r)=f(r), ~~B(r)=\frac{1}{f(r)}, ~~C(r)=r^2,~~f(r)=1-\frac{1}{r}-\frac{S}{r^{\lambda}}.
\end{equation}
From Eq.(\ref{metric3}) we know that the event horizons $r_{H(\lambda)}$  for different $\lambda$ are given by
\begin{eqnarray}
r_{H(2)}&=&\frac{1}{2}(1+\sqrt{1+4S}),\\\label{horizon2}
r_{H(3)}&=&\frac{1}{3}(1+\frac{2^{\frac{1}{3}}}{(2+27S+3\sqrt{3}\sqrt{4S+27S^{2}})^{\frac{1}{3}}}+
\frac{(2+27S+3\sqrt{3}\sqrt{4S+27S^{2}})^{\frac{1}{3}}}{2^{\frac{1}{3}}}),\\\label{horizon3}
r_{H(4)}&=&\frac{1}{4}+\frac{1}{4}\sqrt{1-16L+\frac{4S}{3L}}+\frac{1}{2}\sqrt{\frac{1}{2}+4L-\frac{S}{3L}+
\frac{1}{2\sqrt{1-16L+\frac{4S}{3L}}}} ,\label{horizon4}
\end{eqnarray}
with
\begin{eqnarray}
L=\frac{(\frac{2}{3})^{\frac{1}{3}}S}{(-9S+\sqrt{3}\sqrt{27S^{2}+256S^{3}})^{\frac{1}{3}}}.\nonumber
\end{eqnarray}
The limit for $S$ could be drawn from the above equations: when the event horizon exists, we must confine $S>-0.25$, $S>-0.148$ and $S>-0.105$ for $\lambda=2$, $\lambda=3$ and $\lambda=4$, respectively. Obviously, when $S=0$, the radius of the event horizon will always be $1$ for fixed $\lambda$, the same value as the Schwarzschild case.

The equation of the photon sphere is given by \cite{Vir2,Vir3}
\begin{equation}\label{u1}
\frac{C'(r)}{C(r)}=\frac{A'(r)}{A(r)},
\end{equation}
which admits at least one positive solution and the largest real root of Eq.(\ref{u1}) is defined as the radius of the photon sphere $r_{ps(\lambda)}$. The $r_{ps(\lambda)}$ for $\lambda=2,~3$ and $4$ are given as
\begin{eqnarray}
r_{ps(2)}&=&\frac{1}{4}(3+\sqrt{9+32S}),\\\label{U2}
r_{ps(3)}&=&\frac{1}{2}+\frac{1+(1+10S-2\sqrt{5}\sqrt{S+5S^2})^{\frac{2}{3}}}
{2(1+10S-2\sqrt{5}\sqrt{S+5S^2})^{\frac{1}{3}}}\\\label{u3}
r_{ps(4)}&=&\frac{3}{8}+\frac{1}{8}\sqrt{9-\frac{128S}{K}+8K}+\frac{1}{2}\sqrt{\frac{9}{8}+\frac{8S}{K}-\frac{K}{2}+
\frac{27}{8\sqrt{9-\frac{128S}{K}+8K}}},\label{u4}
\end{eqnarray}
with
\begin{eqnarray}
K=(-279-\sqrt{729S^2+4096S^3})^{\frac{1}{3}}.\nonumber
\end{eqnarray}
\begin{figure}[!htb]
\includegraphics[width=0.96\textwidth ]{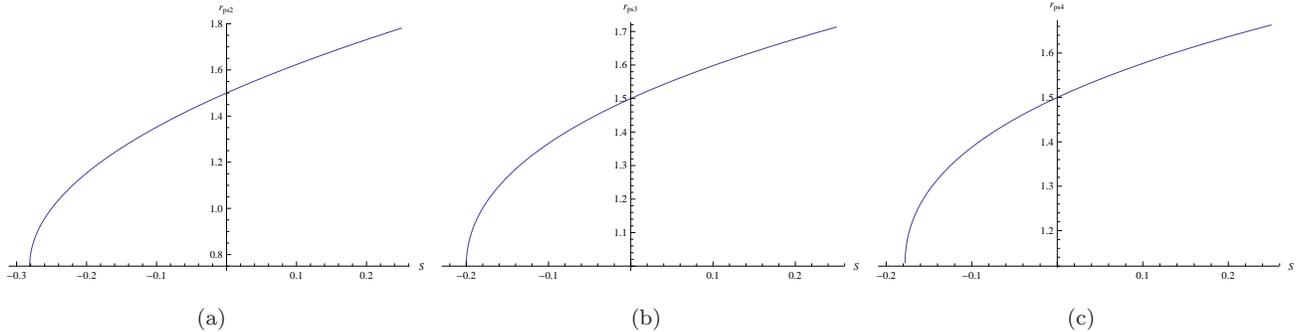}
\caption{(color online) The radius of the photon sphere changes with parameter $S$ for different $\lambda$ in massive gravity.}
\end{figure}
The limit for $S$ could also be drawn from the above equations and figure, which are $S>-0.28$, $S>-0.2$ and $S>-0.1779$ for $\lambda=2$, $\lambda=3$ and $\lambda=4$, respectively. It is obviously that the constraint is lower than that for the event horizon. Therefore, we should use the limit $S>-0.25$, $S>-0.148$ and $S>-0.105$ for $\lambda=2$, $\lambda=3$ and $\lambda=4$. It is not hard to find that the radius of the photon sphere will always be 1.5 for fixed $\lambda$ when $S=0$, the same value as the Schwarzschild case \cite{Bozza}.

Following Ref. \cite{Einstein}, we can define the deflection angle for the photon coming from infinite in the massive gravity as
\begin{equation}\label{angle1}
\alpha(r_{0})=I(r_{0})-\pi,
\end{equation}
where $r_{0}$ is the closest approach distance and $I(r_{0})$ \cite{Vir} is
\begin{equation}\label{angle2}
I(r_{0})=2\int^{\infty}_{r_{0}}\frac{\sqrt{B(r)}dr}{\sqrt{C(r)}\sqrt{\frac{C(r)A(r_{0})}{C(r_{0})A(r)}}-1}.
\end{equation}
The deflection angle increases when parameter $r_{0}$ decreases. For a special value of $r_{0}$ the deflection angle will become $2\pi$, it means that the light ray will make a complete loop around the compact object before reaching the observer. What's more, the deflection angle diverges and the photon is captured when $r_{0}$ reduces until equal to the radius of the photon sphere $r_{ps}$.

In order to find the behavior of the deflection angle very close to the photon sphere, we adopt the evaluation method for the integral (14) proposed by Bozza
\footnotemark[2]\footnotetext[2]{Virbhadra \cite{Vir5} who coined the term relativistic images showed that deflection angles and image positions of
 relativistic images for the Schwarzschild lensing obtained by Bozza \cite{Bozza1} have about $0.5\%$ error.
However, we found that Bozza's more recent analytic formalism \cite{Bozza} has
about $0.38\%$ errors. Therefore, we adopt here Bozza's new (instead of
his old method) formalism \cite{Bozza} to study the deflection angles and image
positions for relativistic images. However, as shown by Virbhadra
\cite{Vir5}, Bozza's method \cite{Bozza2} gives huge errors, about $-31.8\%$ to $20.0\%$,
for time delays among relativistic images. Virbhadra supported these
results by truly convincing physical arguments. Therefore, for time
delays of relativistic images we adopt Virbhadra and Keeton formalism
\cite{Vir4,Vir5}. Virbhadra \cite{Vir5} also showed that Bozza's results have large
errors for magnifications of relativistic images as well.}\cite{Bozza}.
Let us define a variable

\begin{equation}\label{variable}
z=1-\frac{r_{0}}{r},
\end{equation}
then we will obtain
\begin{equation}\label{angle3}
I(r_{0})=\int^{1}_{0}R(z,r_{0})f(z,r_{0})dz,
\end{equation}
where
\begin{equation}\label{R}
R(z,r_{0})=\frac{2r^{2}\sqrt{A(r)B(r)C(r_{0})}}{r_{0}C(r)},\nonumber
\end{equation}
\begin{equation}\label{f}
f(z,r_{0})=\frac{1}{\sqrt{A(r_{0})-\frac{A(r)C(r_{0})}{C(r)}}}.
\end{equation}
Moreover, the function $R(z,r_{0})$ is regular for all values of $z$ and $r_{0}$. While the function $f(z,r_{0})$ diverges as $z$ tends to zero, i.e., the photon approaches the photon sphere. Thus, the integral (\ref{angle3}) can be split into \begin{eqnarray}\label{IDR}
I_{D}(r_{0})&=&\int^{1}_{0}R(0,r_{ps})f_{0}(z,r_{0})dz, \\
I_{R}(r_{0})&=&\int^{1}_{0}[R(z,r_{0})f(z,r_{0})-R(0,r_{ps})f_{0}(z,r_{0})]dz,
\end{eqnarray}
where $I_{D}(r_{0})$ and $I_{R}(r_{0})$ denote the divergent and regular parts in the integral (\ref{angle3}), respectively. For the purpose of finding the order of divergence of the integrand, we expand the argument of the square root in $f(z,r_{0})$ to the second order in $z$, then we have

\begin{equation}\label{f0}
f_{0}(z,r_{0})=\frac{1}{\sqrt{p_{\lambda}(r_{0})z+q_{\lambda}(r_{0})z^{2}}},
\end{equation}
where
\begin{eqnarray}\label{pq}
\nonumber p_{2}(r_{0})&=&2-\frac{3}{r_{0}}-\frac{4S}{r_{0}^{2}},~~~~
q_{2}(r_{0})=-1+\frac{3}{r_{0}}+\frac{6S}{r_{0}^{2}},\\\nonumber
p_{3}(r_{0})&=&2-\frac{3}{r_{0}}-\frac{5S}{r_{0}^{3}},~~~~\nonumber
q_{3}(r_{0})=-1+\frac{3}{r_{0}}+\frac{10S}{r_{0}^{3}}, \\
p_{4}(r_{0})&=&2-\frac{3}{r_{0}}-\frac{6S}{r_{0}^{4}},~~~~
q_{4}(r_{0})=-1+\frac{3}{r_{0}}+\frac{15S}{r_{0}^{4}}.
\end{eqnarray}
Obviously, we can find that the coefficients $p_\lambda(r_{0})$ vanish and the leading term of the divergence in $f_{0}(z,r_{0})$ is $z^{-1}$ when $r_{0}$ is equal to the radius of the photon sphere $r_{ps}$ for different $\lambda$, which means that the integral (\ref{angle3}) diverges logarithmically. Therefore the deflection angle in the strong field region can be expanded in the form
\begin{equation}\label{angle4}
\alpha(\theta)=-\bar{a}\log(\frac{\theta D_{OL}}{u_{ps}}-1)+\bar{b}+o(u-u_{ps}),
\end{equation}
with
\begin{eqnarray}\label{ab}
\nonumber \bar{a}&=&\frac{R(0,r_{ps})}{2\sqrt{q_{\lambda}(r_{ps})}},\\
\nonumber  \bar{b}&=&-\pi+b_{R}+\bar{a}\log\frac{r_{ps}^{2}[C''(r_{ps})A(r_{ps})-C(r_{ps})A''(r_{ps})]}{u_{ps}\sqrt{A^{3}(r_{ps})C(r_{ps})}},\\
\nonumber b_{R}&=&I_{R}(r_{ps}),\\
u_{ps}&=&\frac{r_{ps}}{\sqrt{A(r_{ps})}},
\end{eqnarray}
where $D_{OL}$ is the distance between observer and gravitational lens, $\theta$ is the angular image separation between the optical axis and the image, and they are satisfied $u=\theta D_{OL}$. $u_{ps}$ is the impact parameter $u$ evaluated at $r_{ps}$, called minimum impact parameter. $\bar{a}$ and $\bar{b}$ are the so-called strong field limit coefficients which depend on the metric functions evaluated at $r_{ps}$ and they are the theoretical predictions, not the observables. Making use of Eqs.(\ref{angle4}) and (\ref{ab}), we can study the properties of strong gravitational lensing in the massive gravity.

\begin{figure}[htbp]
\begin{center}
\includegraphics[scale=1.]{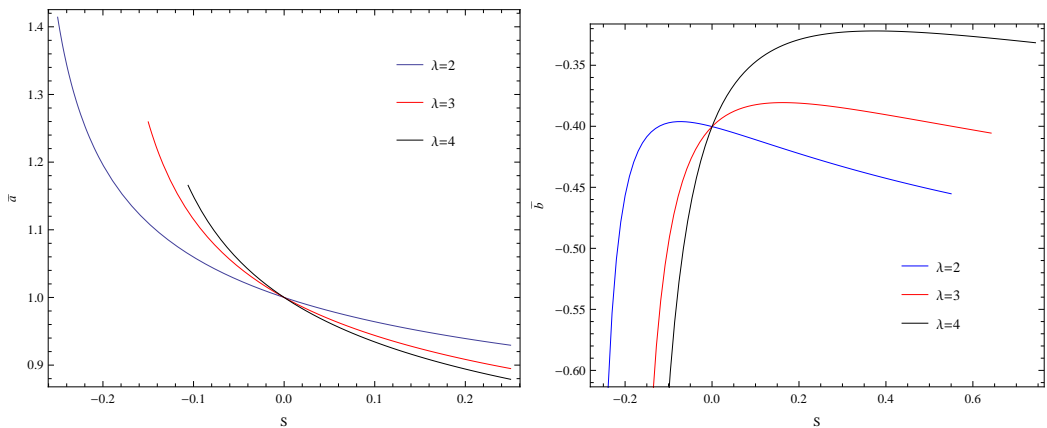}
\end{center}
\caption{(color online)  Variation of the strong field limit coefficients $\bar{a}$ (left) and $\bar{b}$ (right) with parameters $S$ and $\lambda$ in the massive gravity. The intersection of $S=0$ stands for Schwarzschild case.}
\label{ab1}
\end{figure}

\begin{figure}[htbp]
\begin{center}
\includegraphics[scale=1.]{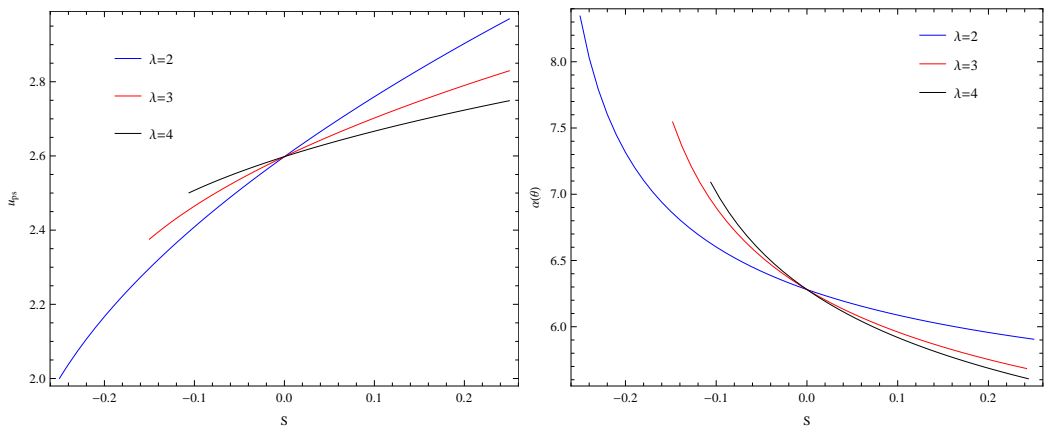}
\end{center}
\caption{(color online)  The left one is the variation of the minimum impact parameter $u_{ps}$ with parameters $S$ and $\lambda$ in the massive gravity. The right one is the deflection angle evaluated at $u=u_{ps}+0.003$ as function of $S$ for different $\lambda$ in massive gravity. The intersection of $S=0$ stands for Schwarzschild case.}
\label{u ar}
\end{figure}

In Fig.(\ref{ab1}), we plot the change of the coefficients $\bar{a}$ and $\bar{b}$ with scaler charge $S$ for different $\lambda$ in the massive gravity. We can see,  with the increase of $S$ for fixed $\lambda$,  that the coefficient $\bar{a}$ decreases,  but $\bar{b}$ increases first, after reaching a maximum, it decreases slowly. Moreover, $\bar{a}$ increases with the increase of $\lambda$ for $S<0$ but decreases for $S>0$. Nonetheless, the variation of $\bar{b}$ with $\lambda$ is converse to the variation of $\bar{a}$ with $\lambda$. The Fig.(\ref{u ar}) tells us that the minimum impact parameter $u_{ps}$ grows with  the increase of $S$ for fixed $\lambda$, and it increases with the increase of $\lambda$ if $S<0$ but decreases if $S>0$. We present the deflection angle $\alpha(\theta)$ that evaluated at $u=u_{ps}+0.003$ in Fig.(\ref{u ar}), too, which shows that in the strong field limit the deflection angle has the similar properties with the coefficient $\bar{a}$. This means that the deflection angle of the light rays is dominated by the logarithmic term in the strong gravitational lensing. From all of the above figures, we can see clearly that all parameters have different properties when they vary with $\lambda$ for $S<0$ and $S>0$. This is due to that the different values of scalar charge $S$ lead to different physical effects. When $S>0$, the gravitational potential is attractive and stronger than that of the Schwarzschild black hole.  When $S<0$, even though the Newton potential is always attractive form the event horizon to large distances, the attraction is weaker than that of the Schwarzschild black hole. Furthermore, it is not hard to find that for any $\lambda$, each line of $\bar{a}$, $\bar{b}$, $u_{ps}$ and $\alpha(\theta)$ intersects at $S=0$, which means that they recover the results of the standard Schwarzschild case, i.e.,  $\bar{a}=1$, $\bar{b}=-0.4002$, $u_{ps}=2.598$ and  $\alpha(\theta)=6.28$ \cite{Bozza}. We should note that the metric is the same as the Reissner-Nordstr$\ddot{o}$m metric when $\lambda=2$ and $S<0$, therefore the blue line of Figs. (\ref{ab1}) and (\ref{u ar}) in the region of $\lambda=2$ and $S<0$ is the same as the results for Reissner-Nordstr$\ddot{o}$m black hole \cite{Bozza,Eirc}.

\section{Observables in the strong deflection limit}

Now let us see how the parameters $S$ and $\lambda$ affect the observables in the strong gravitational lensing. Considering the source and observer are far enough from the lens, and the source, lens and observer are highly aligned. Then the lens equation can be written as \cite{Bozza}
\begin{equation}\label{gamma}
\beta=\theta-\frac{D_{LS}}{D_{OS}}\triangle\alpha_{n},
\end{equation}
where $D_{LS}$ is the distance between the lens and the source, $D_{OS}$ is the distance between the observer and the source and $D_{OS}=D_{LS}+D_{OL}$. $\beta$ is the angle between the direction of the source and the optical axis, called angular source position. It is well-known that a light ray can pass close to the photon sphere and go around the lens once, twice, thrice, or many times  depending on the impact parameter $u$ before reaching the observer. Therefore, we take $\triangle\alpha_{n}=\alpha-2n\pi$ as the offset of deflection angle, and $n$ represents the loop numbers of the light turns. Thus, a massive compact lens gives rise an infinite sequence of images on both sides of the optic axis. Virbhadra and Ellis in their PRD paper\cite{Vir1} called these images which are formed due to the bending of light through more than $\frac{3}{2}\pi$ relativistic images, as the light rays giving rise to them pass through a strong gravitational field before reaching the observer.

We can find that the angular separation between the lens and the $n-th$ relativistic image is
\begin{equation}\label{theta}
\theta_{n}\simeq\theta^{0}_{n}+\frac{u_{ps}e_{n}
(\beta-\theta_{n}^{0})D_{OS}}{\bar{a}D_{LS}D_{OL}},
\end{equation}
with
\begin{equation}\label{theta1}
\theta_{n}^{0}=\frac{u_{ps}}{D_{OL}}(1+e_{n}),~~~~
e_{n}=e^{\frac{\bar{b}-2n\pi}{\bar{a}}},
\end{equation}
where the quantity $\theta_{n}^{0}$ is the angular image position corresponding to $\alpha=2n\pi$.

The magnification of the $n-th$ relativistic image is given by

\begin{equation}\label{magnification}
\mu_{n}=\frac{1}{\frac{\beta}{\theta}\frac{\partial\beta}{\partial\theta}}\mid_{\theta^{0}_{n}}
=\frac{u_{ps}^{2}e_{n}(1+e_{n})D_{OS}}{\bar{a}\beta D^{2}_{OL}D_{LS}}.
\end{equation}

Obviously, the first relativistic image is the brightest, and the magnification decreases exponentially with $n$. Hence we separate at least the outermost and brightest image $\theta_{1}$ from all the others which are packed together at $\theta_{\infty}$ \cite{Bozza,Bozza2}. As $n\rightarrow\infty$, we can find that $e_{n}\rightarrow0$ from Eq. (\ref{theta1}), which implies that the minimum impact parameter $u_{ps}$ and the asymptotic position of a set of images $\theta_{\infty}$ obey a simple form
\begin{equation}\label{theta2}
u_{ps}=D_{OL}\theta_{\infty}.
\end{equation}
Thus the angular image separation $s$ between the first image and the other ones, the ratio $\mathcal{R}$ of the flux from the first image to those from all other images can be expressed as
\begin{eqnarray}\label{s r}
 \nonumber   s&=&\theta_{1}-\theta_{\infty}=\theta_{\infty} e^{\frac{\bar{b}-2\pi}{\bar{a}}},\\
\mathcal{R}&=&\frac{\mu_{1}}{\Sigma^{\infty}_{n=2}\mu_{n}}=e^{\frac{2\pi}{\bar{a}}}.
\end{eqnarray}
These two formulas can be easily inverted to give
\begin{eqnarray}\label{ab2}
\nonumber \bar{a}&=&\frac{2\pi}{\log \mathcal{R}},\\
\bar{b}&=&\bar{a}\log(\frac{\mathcal{R}s}{\theta_{\infty}}).
\end{eqnarray}
Through measuring $s$, $\theta_{\infty}$ and $\mathcal{R}$, we can obtain the strong deflection limit coefficients $\bar{a}$ and $\bar{b}$ and the minimum impact parameter $u_{ps}$. Comparing their values with those predicted by theoretical models, we can obtain information about the parameters of the lens object that stored in them.

\begin{figure}[htbp]
\begin{center}
\includegraphics[scale=1.]{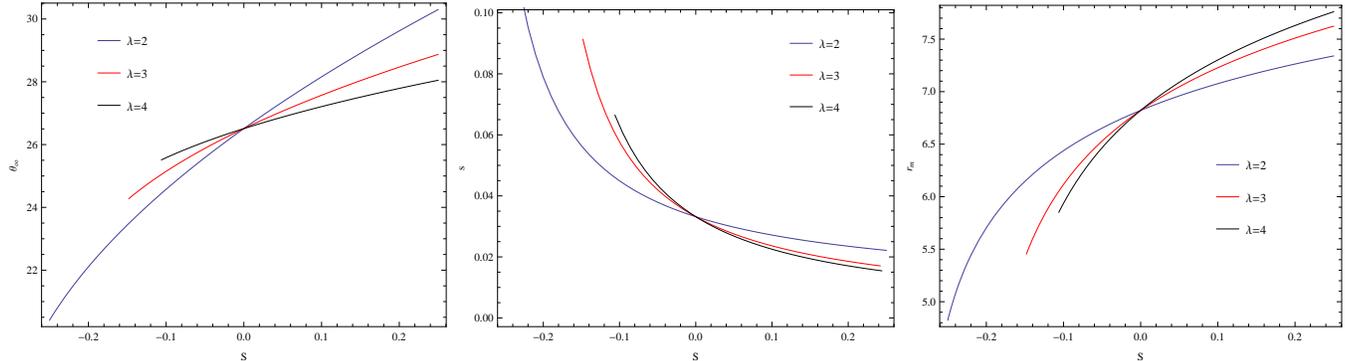}
\end{center}
\caption{(color online) Gravitational lensing by the galactic center black hole. Variation of the values of the angular image position $\theta_{\infty}$, the angular image separation $s$ and relative magnitudes $r_{m}$ with parameters $S$ and $\lambda$. All angles are expressed in \emph{$\mu$arcsec}. The intersection of $S=0$ stands for Schwarzschild case.}
\label{thsrm}
\end{figure}

Suppose that there is  a  supermassive black hole in galactic center \cite{Genzel} which has a mass $M=4.4\times10^{6}M_{\odot}$ and is situated at a distance from the Earth $D_{OL}= 8.5kpc$, then the ratio of the mass to the distance is $\frac{M}{D_{OL}}\approx2.4734\times10^{-11}$. Take advantage of Eqs.(\ref{angle4}), (\ref{theta2}) and (\ref{s r}) we can estimate the values of the coefficients and observation parameters for gravitational lensing in the strong field limit. The numerical value for the angular image position $\theta_{\infty}$, the angular image separation $s$ and the relative magnifications $r_{m}$ (which is related to $\mathcal{R}$ by $r_{m}=2.5\log \mathcal{R}$) of the relativistic images are listed in Table I, and then plotted in Fig.(\ref{thsrm}).

\begin{table}[htbp]
\begin{center}
\begin{tabular}{|c|c|c|c|c|c|c|c|c|c|c|}
\hline
\hline
&\multicolumn{3}{c|}{$\theta_{\infty}$($\mu$arcsec)}
&\multicolumn{3}{c|}{$s$($\mu$ arcsec)}
&\multicolumn{3}{c|}{$r_{m}$(magnitude)} \\
\hline
$S$&$\lambda=2$&$\lambda=3$&$\lambda=4$&$\lambda=2$&$\lambda=3$&
$\lambda=4$&$\lambda=2$&$\lambda=3$&$\lambda=4$\\
\hline
 -0.10& 24.58&25.15&25.58&0.0449&0.0577& 0.0629&6.44&6.12&5.94\\
 \hline
-0.05&25.59&25.88&26.09&0.0379&0.0420&0.0435&6.65&6.53&6.47\\
\hline
0& 26.51&26.51&26.51&0.0332&0.0332&0.0332&6.82&6.82&6.82 \\
\hline
0.05& 27.36&27.07&26.88&0.0298&0.0275&0.0268&6.96&7.05&7.09\\
 \hline
0.10&28.16&27.57&27.21&0.0272&0.0236&0.0225&7.08&7.23&7.30\\
\hline
\hline
\end{tabular}
\end{center}
\label{tab1} \caption{Numerical estimation for main observation parameters in
the strong field limit for the black hole at the center of our galaxy, which is supposed to be described by the black hole with scalar charge in massive gravity.}
\end{table}

Clearly, we can find that for fixed $\lambda$ with the increase of $S$, the angular image position $\theta_{\infty}$ and the relative magnifications $r_{m}$ of the relativistic images increase, while the angular image separation $s$ decreases. What's more, as $S$ and $\lambda$ change, the variation of the angular image position $\theta_{\infty}$ of the relativistic images is similar to the minimum impact parameter $u_{ps}$, and the angular image separation $s$ is similar to the deflection angle $\alpha(\theta)$. Then, the relative magnifications $r_{m}$ increases as the parameter $\lambda$ increases for  $S>0$, but it changes contrary when $S<0$. From Fig.(\ref{thsrm}), it is interesting to find that for different $\lambda$, each line of $\theta_{\infty}$, $s$ and $r_{m}$ intersects at $S=0$, which returns standard Schwarzschild case and $\theta_{\infty}=26.5095\mu as$, $s=0.0331\mu as$ and $r_{m}=6.82$ \cite{Bozza}. The blue line of Fig.(\ref{thsrm}) in the regional of $S<0$, $\lambda=2$ gives result for the Reissner-Nordstr$\ddot{o}$m black hole \cite{Bozza,Eirc}.

\section{Time delay in massive gravity}
It is well known that the time delay of light traveling from the source to the observer with the closest distance of approach $r_{0}$ is defined as the difference between the light travel time for the actual ray in the gravitational field of the lens (deflector) and the travel time for the straight path between the source and the observer in the absence of the lens (i.e., if there were no gravitational fields).  Following the method used in \cite{Weinberg}, Virbhadra and Keeton \cite{Vir4} first obtain the time required for light to go from one point with $\{r, \theta=\frac{\pi}{2}, \varphi=\varphi_{1}\}$, to a second point $\{r_{0}, \theta=\frac{\pi}{2}, \varphi=\varphi_{2}\}$ as

\begin{equation}\label{time1}
t(r,r_{0})=t(r_{0},r)=\int^{r}_{r_{0}}\sqrt{\frac{A(r)/B(r)}{1-
\frac{B(r)}{B(r_{0})}(\frac{C(r_{0})}{C(r)})^{2}}}dr.
\end{equation}

Using this result, time delay of images in massive gravity can be expressed as \cite{Vir5}

\begin{equation}\label{time2}
\tau(r_{0})=2M[\int^{\chi_{s}}_{r_{0}}\frac{dr}{F(r)}+
\int^{\chi_{o}}_{r_{0}}\frac{dr}{F(r)}]-D_{OS}~sec \beta,
\end{equation}
with
\begin{eqnarray}\label{time4}
F(r)=f(r)\sqrt{1-\frac{f(r)}{f(r_{0})}(\frac{r_{0}}{r})^{2}},~~~~
\chi_{s}=\frac{D_{OS}}{2M}\sqrt{(\frac{D_{LS}}{D_{OS}})^{2}+\tan^{2}\beta},~~~~
\chi_{o}=\frac{D_{OL}}{2M},
\end{eqnarray}
where the first and second terms with positive sign are, respectively, the travel time of the light from the source to the point of closest approach and from that point to the observer, and the third term with a minus sign is the light travel time from the source to the observer in the absence of any gravitational field.

In this part, we just study the primary relativistic image which is on the same side of the source and doesn't loop around the lens ($n=0$). Now, we still use the same mass and distance as the above part to estimate time delay, but with an additional assumption, $D_{OL}=D_{LS}=\frac{1}{2}D_{OS}$. The angular image position $\theta_{sch}$, deflection angle $\alpha_{sch}$ and time delay $\tau_{sch}$ for $S=0$ which is well known as the Schwarzschild case have been shown in Table II. The similar calculation has been done in \cite{Vir4,Vir5} with different mass and distance. After that, we compare the value of $\tau_{\lambda}$ ($\lambda=2,3,4$) with $\tau_{sch}$, and then show the results in Table III. You may ask, why we only calculate the time delay $\tau$ for $\lambda=2,3,4$, this is because that the time delay is easier to be observed than the angular image position and deflection angle.

We can find that the angular image position increases but the deflection angle and time delay decrease with the increase of angular source position for Schwarzschild black hole, from Table II. This is consequence of that with the increase of angular source position $\beta$, the closest distance of approach $r_{0}$ increases, then the effect of the lens will be weakened, and its effect on the light will be reduced. It is easy to find from Table III that with the increase of $\lambda$, the differential time delay $\Delta\tau=\tau_{\lambda}-\tau_{sch}$ decreases significantly, almost about $10^{-5}$ minutes. This is because that the black hole with scalar charge in massive gravity is more close to the Schwarzschild black hole when $\lambda$ is larger. What's more, for each fixed $\lambda$, the $\Delta\tau$ is symmetry for the symmetrical scalar charge. That is to say, the positive scalar charge increases the time delay, and negative scalar charge decreases the time delay, but the magnitude is the same. And the influence of $S=\pm0.1$ is nearly two times of $S=\pm0.05$ for fixed $\lambda$. Furthermore, for each fixed scalar charge $S$ and parameter $\lambda$, $\Delta\tau$ decreases with the increase of angular source position $\beta$. This is the same reason as the Schwarzschild case that the lens is weakened.
\begin{table}[ht]
\begin{center}
\begin{tabular}{|c|c|c|c|}
\hline
\hline
$\beta$&$\theta_{sch}$&$\alpha_{sch}$&$\tau_{sch}$\\
\hline
$0$&1.4512174&2.9024348&18.11567180388526151721\\
\hline
$10^{-6}$&1.4512179&2.9024338&18.11567080981492372465\\
\hline
$10^{-5}$&1.4512224&2.9024248&18.11566186319729594447\\
\hline
$10^{-4}$&1.4512674&2.9023348&18.11557239854682350187\\
\hline
$10^{-3}$&1.4517174&2.9014349&18.11467790460512913207\\
\hline
$10^{-2}$&1.4562259&2.8924519&18.10574820398524547348\\
\hline
$10^{-1}$&1.5020782&2.8041564&18.01795757500632361835\\
\hline
$1$&2.0349350&2.0698701&17.27351876798165832134\\
\hline
$2$&2.7623908&1.5247817&16.66483944815783166894\\
\hline
$3$&3.5871078&1.1742156&16.20694418018656670937\\
\hline
$4$&4.4710356&0.9420713&15.84711090705742086357\\
\hline
$5$&5.3906774&0.7813548&15.55355445096342729969\\
\hline
\hline
\end{tabular}
\end{center}
\label{tab2} \caption{Numerical estimation for angular image position $\theta$, deflection angle $\alpha$, time delay $\tau$ for the black hole at the center of our galaxy, which is supposed to be described by Schwarzschild black hole. $\beta$ stands for the angular source position. Here, all angles are expressed in \emph{arcsec} and time delay is expressed in \emph{minutes}.}
\end{table}

\begin{table}[htbp]
\begin{center}
\begin{tabular}{|c|c|c|c|c|c|c|}
\hline
\hline
&\multicolumn{2}{c|}{$\tau_{2}-\tau_{sch}$($10^{-7}$)}
&\multicolumn{2}{c|}{$\tau_{3}-\tau_{sch}$($10^{-13}$)}
&\multicolumn{2}{c|}{$\tau_{4}-\tau_{sch}$($10^{-18}$)} \\
\hline
$\beta$&$S=\pm0.05$&$S=\pm0.1$&$S=\pm0.05$&$S=\pm0.1$&$S=\pm0.05$&$S=\pm0.1$\\
\hline
0&$\pm$5.9&$\pm$11.9&$\pm$23.8&$\pm$47.6&$\pm$12.3&$\pm$24.6\\
\hline
$10^{-6}$&$\pm$5.9&$\pm$11.9&$\pm$23.8&$\pm$47.6&$\pm$12.3&$\pm$24.6\\
\hline
$10^{-5}$&$\pm$5.9&$\pm$11.9&$\pm$23.8&$\pm$47.6&$\pm$12.3&$\pm$24.6\\
\hline
$10^{-4}$&$\pm$5.9&$\pm$11.9&$\pm$23.8&$\pm$47.6&$\pm$12.3&$\pm$24.6\\
 \hline
$10^{-3}$&$\pm$5.9&$\pm$11.9&$\pm$23.8&$\pm$47.6&$\pm$12.3&$\pm$24.6\\
\hline
$10^{-2}$&$\pm$5.9&$\pm$11.9&$\pm$23.6&$\pm$47.3&$\pm$12.2&$\pm$24.4\\
\hline
$10^{-1}$&$\pm$5.7&$\pm$11.7&$\pm$22.2&$\pm$44.4&$\pm$11.1&$\pm$22.2\\
\hline
1&$\pm$4.2&$\pm$8.5&$\pm$12.1&$\pm$24.2&$\pm$4.4&$\pm$8.9\\
\hline
2&$\pm$3.1&$\pm$6.2&$\pm$6.5&$\pm$13.1&$\pm$1.7&$\pm$3.5\\
\hline
3&$\pm$2.4&$\pm$4.8&$\pm$3.8&$\pm$7.7&$\pm$0.8&$\pm$1.6\\
\hline
4&$\pm$1.9&$\pm$3.8&$\pm$2.5&$\pm$5.0&$\pm$0.4&$\pm$0.8\\
\hline
5&$\pm$1.6&$\pm$3.2&$\pm$1.7&$\pm$3.4&$\pm$0.2&$\pm$0.4\\
\hline
\hline
\end{tabular}
\label{tab3} \caption{Numerical estimation for differential time delay $\Delta\tau=\tau_{\lambda}-\tau_{sch}$ for the black hole at the center of our galaxy, which is supposed to be described by the black hole with scalar charge in massive gravity. Here, the angular source position $\beta$ is expressed in \emph{arcsec} and differential time delay $\Delta\tau$ is expressed in \emph{minutes}.}
\end{center}
\end{table}

\section{Summary}

We investigated the strong gravitational lensing for the black hole with scalar charge in massive gravity. We determined the value range of scalar charge $S$ for fixed $\lambda$ by the existing region of the event horizon at first. Then we found that the scalar charge $S$ and parameter of model $\lambda$ of black hole affect the event horizon $r_{H}$, radius of the photon sphere $r_{ps}$, deflection angle $\alpha(\theta)$, strong field limit coefficients $\bar{a}$ and $\bar{b}$, minimum impact parameter $u_{ps}$ and main observables, such as the angular image position $\theta_{\infty}$, the angular image separation $s$ and the relative magnifications $r_{m}$ of relativistic images in strong gravitational lensing. We also got the time delay by numerical calculation in which we do not take either weak or strong field approximation. The coefficient $\bar{a}$, deflection angle $\alpha(\theta)$ and angular image separation $s$ decrease with the increase of $S$ for fixed $\lambda$, and they increase with the increase of $\lambda$ for $S<0$ but decrease with the increase of $\lambda$ for $S>0$. The coefficient $\bar{b}$ is a little special, with the increase of $S$ for fixed $\lambda$, it increases first, then decreases slowly after reaching a maximum. And the variation of $\bar{b}$ with $\lambda$ is converse to the variation of $\bar{a}$ with $\lambda$. What's more, there are two other parameters, minimum impact parameter $u_{ps}$ and angular image position $\theta_{\infty}$ of the relativistic images, that change similar too, i.e., they rise with the increase of $S$ for different $\lambda$ but they increase for $S<0$ and decrease for $S>0$, with the increase of $\lambda$. Then, the relative magnifications $r_{m}$ increases with the increase of $S$ for fixed $\lambda$ while it grows with $\lambda$ when $S>0$ and changes contrary when $S<0$. The reason to why all parameters have different properties when they vary with $\lambda$ for $S<0$ and $S>0$ is that different scalar charge $S$ leads to different physical effects. In $S>0$ case, the gravitational potential is attractive and stronger than that of the Schwarzschild black hole. In $S<0$ case, even though the Newton potential is always attractive form the event horizon to large distance, the attraction is weaker than that of the Schwarzschild black hole. The time delay decreases with the increase of angular source position $\beta$, because of that with the increase of $\beta$, the closest distance of approach increases, then the effect of the lens will be weakened, and its effect on the light will be reduced. Due to the black hole with scalar charge in massive gravity is more close to the Schwarzschild case when $\lambda$ is larger, the differential time delay $\tau_{\lambda}-\tau_{sch}$ decreases remarkable with the increase of $\lambda$.
And the influence of scalar charge $S$ on time delay is very small but regular compared with $\beta$ and $\lambda$. It is should point that this black hole and the results will recover to Schwarzschild case at $S=0$, i.e., $r_{H}=1$, $r_{ps}=1.5$, $\bar{a}=1$, $\bar{b}=-0.4002$, $u_{ps}=2.598$, $\alpha(\theta)=6.28$, $\theta_{\infty}=26.5095\mu as$, $s=0.0331\mu as$, $r_{m}=6.82$, and the black hole taking the same form as the Reissner-Nordstr$\ddot{o}$m case when $\lambda=2$ and $S<0$, so the blue line of every figure in the regional of $\lambda=2$, $S<0$ gives result for the Reissner-Nordstr$\ddot{o}$m black hole.

\begin{acknowledgments}
{{We thank Xiaokai He and Xiongjun Fang for suggestions and numerical calculation. This work is supported by the  National Natural Science Foundation
of China under Grant Nos. 11475061 and 11305058;  the SRFDP under Grant No.
20114306110003; the Open Project Program of State Key Laboratory of
Theoretical Physics, Institute of Theoretical Physics, Chinese
Academy of Sciences, China (No.Y5KF161CJ1); the Hunan Provincial Innovation Foundation for postgraduate (Grant No.CX2016B164).}}

\end{acknowledgments}

\end{document}